\documentclass[showpacs,amsmath,amssymb,prd]{revtex4}
\usepackage{graphicx}% Include figure files
\usepackage{dcolumn}% Align table columns on decimal point
\usepackage{bm}% bold math
\usepackage{epsfig}% bold math

\newcommand{\be}{\begin{enumerate}}
\newcommand{\ee}{\end{enumerate}}

\begin{document}

\title{\boldmath Observation of $K^*(892)^0\bar K^*(892)^0$ in
$\chi_{cJ}$ Decays}

%\vspace*{10pt}
\author{M.~Ablikim$^{1}$, J.~Z.~Bai$^{1}$, Y.~Ban$^{10}$,
J.~G.~Bian$^{1}$, X.~Cai$^{1}$, J.~F.~Chang$^{1}$,
H.~F.~Chen$^{16}$, H.~S.~Chen$^{1}$, H.~X.~Chen$^{1}$,
J.~C.~Chen$^{1}$
% ...
\vspace{0.2cm}
\\(BES Collaboration)\\
\vspace{0.1cm}
{\small\it
$^1$ Institute of High Energy Physics, Beijing 100039, People's
Republic of China\\
% ...
$^{10}$ Peking University, Beijing 100871, People's Republic of
China\\
% ...
$^{16}$ University of Science and Technology of China, Hefei 230026,
People's Republic of China\\
% ...
}
}
%\vspace*{0.4cm}
\date{\today}

\small
%\vspace*{0.1cm}
\begin{abstract}
$K^*(892)^0\bar K^*(892)^0$ signals from $\chi_{cJ}~(J=0,1,2)$ decays
are observed for the first time using a data sample of 14 million
$\psi(2S)$ events accumulated at the BES\,II detector.  The branching
fractions ${\cal B}(\chi_{cJ}\to K^*(892)^0\bar K^*(892)^0)$~$(J =
0,1,2)$ are determined to be $(1.55 \pm 0.35 \pm 0.30)\times 10^{-3}$,
$(1.58 \pm 0.32 \pm 0.29 )\times 10^{-3}$, and $(4.67 \pm 0.55 \pm
0.85 )\times 10^{-3}$ for the $\chi_{c0}$, $\chi_{c1}$ and $\chi_{c2}$
decays, respectively, where the first errors are statistical and the
second are systematic.  The significances of these signals are about
4.2$\sigma$, 4.3$\sigma$, and 7.5$\sigma$, respectively.
\end{abstract}

%%%%\vspace{0.4cm}

\normalsize

%\large
\pacs{13.25.Gv, 12.38.Qk, 14.40.Gx}
%{\flushleft PACS: {13.25.Gv, 14.20.Jn}}% PACS, the Physics and Astronomy}
\maketitle

%\section{\label{sec:level1}First-level heading:\protect\\ The line
%break was forced \lowercase{via} \textbackslash\textbackslash}

%\newpage

%\vspace*{1pt}

%\newpage
\section{Introduction}

Exclusive quarkonium decays constitute an interesting laboratory for
investigating perturbative quantum chromodynamics (QCD). In the case
of $P$-wave charmonium $\chi_{cJ}$ decays to a pair of pseudoscalars,
one finds that the lowest Fock state, the color-singlet contribution, alone
is not sufficient to accommodate the data. Indeed, it turns out that
the color-octet contribution from the next higher Fock state
contributes at the same level as the color singlet one. Its inclusion
yields good agreement with experimental data [1,2].  The calculation
of the partial width of $\chi_{cJ}\to p\bar p$, taking into account
the color octet mechanism \cite{T4}, also obtains results in
reasonable agreement with measurements~\cite{PDG}.  Nevertheless a
recent measurement of the $\chi_{cJ}\to \Lambda\bar \Lambda$ \cite{L}
only agrees marginally with this prediction.

At present there are no predictions for the majority of the hadronic
decay modes.  In addition, few two-body decays have been measured. A
consistent set of predictions for the branching fractions, as well as
more precise experimental measurements, for a number of the two-body
decays may lead to further insight into the nature of these $^3P_J$
$c\bar c$ bound states.

In this paper, we report on the analysis of $\pi^+\pi^-K^+K^-$
final states from $\chi_{cJ}~(J=0,1,2)$ decays using
14 million $\psi(2S)$ events accumulated at the upgraded BES detector
(BES\,II). Signals of $\chi_{c0}$, $\chi_{c1}$ and $\chi_{c2}$ decays
to $K^*(892)^0\bar K^*(892)^0$ in $\psi(2S)$ radiative decays
are observed for the first time.

%\vspace*{1pt}
\section{Bes detector}

BES\,II is a large
solid-angle magnetic spectrometer that is described in detail in Ref.
\cite{BESII}. Charged particle momenta are determined with a
resolution of $\sigma_p/p = 1.78\%\sqrt{1+p^2}$~($p$ in GeV$/c$) in a
40-layer cylindrical drift chamber. Particle identification is
accomplished by specific ionization ($dE/dx$) measurements in the
drift chamber and time-of-flight (TOF) measurements in a barrel-like
array of 48 scintillation counters. The $dE/dx$ resolution is
$\sigma_{dE/dx} = 8.0\%$; the TOF resolution is $\sigma_{TOF} = 180$
ps for Bhabha events. Outside of the time-of-flight counters is a
12-radiation-length barrel shower counter (BSC) comprised of gas
proportional tubes interleaved with lead sheets. The BSC measures the
energies of photons with a resolution of
$\sigma_E/E\simeq 21\%/\sqrt{E}$~($E$ in GeV). Outside the solenoidal
coil, which provides a 0.4 T magnetic field over the tracking
volume, is an iron flux return that is instrumented with three double layers of counters that are
used to identify muons.

In this analysis, a
 GEANT3 based Monte Carlo simulation package (SIMBES) with detailed
   consideration of detector performance (such as dead
   electronic channels) is used.
    The consistency between data and Monte Carlo has been
 checked in
   many high purity physics channels, and the agreement is quite
 reasonable.

\section{Event selection}

The selection criteria described below are similar to those used in a
previous BES analysis \cite{BESc}.

\subsection{Photon identification}

A neutral cluster is considered to be a photon candidate when the
angle between the nearest charged track and the cluster is greater
than 15$^{\circ}$, the first hit is in the beginning six radiation
lengths, and the difference between the angle of the cluster
development direction in the BSC and the photon emission direction is
less than 30$^{\circ}$. The photon candidate with the largest energy
deposit in the BSC is treated as the photon radiated from $\psi(2S)$
and used in a four-constraint kinematic fit to the hypothesis
$\psi(2S)\to\gamma\pi^+\pi^-K^+K^-$.

\subsection{Charged particle identification}

%\vspace*{-5pt}
Each charged track, reconstructed using the MDC information, is
required to be well fit to a three-dimensional helix, be in the polar
angle region $|\cos\theta_{{MDC}}| < 0.80$, and have the point of
closest approach of the track to the beam axis be within 2 cm
of the beam axis and within 20 cm from the center of the interaction
region along the beam line. For each track, the TOF and $dE/dx$
measurements are used to calculate $\chi^2$ values and the
corresponding confidence levels for the hypotheses that the particle is
a pion, kaon or proton (Prob$_{\pi}$, Prob$_{K}$,
Prob$_{p}$).

\subsection{Event selection criteria}

\vspace*{-5pt}
Candidate events are
required to satisfy the following selection criteria:

(1) The number of charged tracks is required to be four with net
charge zero.

%(2) The maximum number of neutral clusters in an event is
%ten.

(2) The sum of the momenta of the two lowest momentum
tracks is required to be greater than 650 MeV; this removes
contamination from $\psi(2S)\to\pi^+\pi^- J/\psi$ events and some 
           of the $\rho^0\pi\pi$ background.

(3) The $\chi^2$ probability for the four-constraint kinematic fit to
the decay hypothesis $\psi(2S)\to\gamma\pi^+\pi^-K^+K^-$ is required
to be greater than 0.01.

A combined probability determined from the four-constraint kinematic
fit and particle identification information is used to separate
$\gamma\pi^+\pi^-\pi^+\pi^-$, $\gamma K^+ K^- K^+ K^-$, and the
different possible particle assignments for the
$\gamma\pi^+\pi^-K^+K^-$ final states. This combined probability,
Prob$_{all}$, is defined as $$\mbox{Prob}_{all} =
\mbox{Prob}(\chi^2_{all}, \mbox{ndf}_{all}),$$ where $\chi^2_{all}$ is
the sum of the $\chi^2$ values from the four-constraint kinematic fit
and those from each of the four particle identification assignments,
and ndf$_{all}$ is the corresponding total number of degrees of
freedom. For an event to be selected,
Prob$_{all}$ of the $\gamma\pi^+\pi^-K^+K^-$ must be larger than those
of the other possibilities.  In
addition, the particle identification probability of each charged
track Prob$_{ID}$ must be $>$ 0.01.

The invariant mass distribution for the $\pi^+\pi^-K^+K^-$ events
that survive all the selection requirements is shown in Fig. 1. There
are clear peaks corresponding to the $\chi_{cJ}$ states. The highest
mass peak
corresponds to charged tracks final states that are kinematically
fit with an unassociated, low energy photon.

\begin{figure}[hbtp]
\begin{center}
\epsfxsize=6.25cm\epsffile{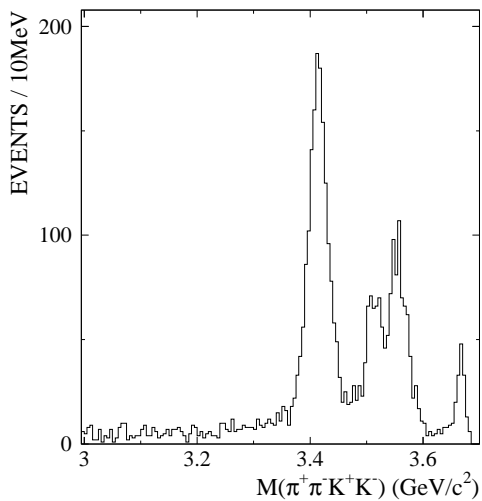}
%\hspace*{0.5cm}
%\epsfxsize=5.250cm\epsffile{mksic-psip.epsi}
\label{cpkp}
%\vspace*{-15pt}
\caption{The $\pi^+\pi^-K^+K^-$ invariant mass spectrum.  }
\end{center}
\end{figure}

\section{ANALYSIS RESULTS}

The scatter plots of $K^-\pi^+$ versus $K^+\pi^-$
invariant
masses for events with a $\pi^+\pi^-K^+K^-$ mass within
(3.30, 3.48) GeV, (3.48, 3.53) GeV and (3.53, 3.65) GeV are shown in
Fig. 2.  Clear $K^*(892)^0\bar K^*(892)^0$
signals can be seen in
all $\chi_{cJ}$ decays, as well as
some hints of $K^*_2(1430)K^*_2(1430)$ (or
$K^*_0(1430)K^*_0(1430)$) and $K_1(1270)K$ (or $K_1(1400)K$)
signals.
In this paper, we study
$K^*(892)^0\bar K^*(892)^0$ production in $\chi_{c0,1,2}$ decays.

\vspace*{5pt}

\begin{figure}[htbp]
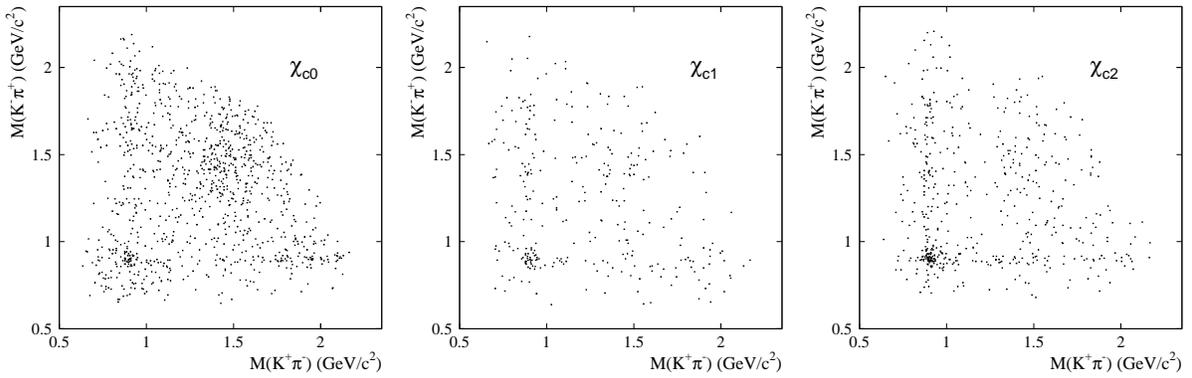

\begin{center}
\epsfxsize=5.00100cm\epsffile{scak-chic0.epsi}
\hspace*{0.1cm}
\epsfxsize=5.00100cm\epsffile{scak-chic1.epsi}
%\vspace*{0.15cm}
\hspace*{0.1cm}
\epsfxsize=5.0010cm\epsffile{scak-chic2.epsi}
%\vspace*{-10pt}
\label{cpkp1}
\caption{Scatter plots of $K^-\pi^+$ versus $K^+\pi^-$ invariant
masses for selected $\gamma\pi^+\pi^-K^+K^-$ events with
$\pi^+\pi^-K^+K^-$ mass in $\chi_{c0}$, $\chi_{c1}$, and $\chi_{c2}$
 mass regions, respectively.  }
\end{center}
\end{figure}

\subsection{$K^*(892)^0\bar K^*(892)^0$ signal}

For the events in $\chi_{cJ}$ mass region (3.30, 3.65) GeV, after
requiring that the mass of either (or both) $K\pi$ pair lies between 0.836
and 0.956 GeV, the mass distribution of the other $K\pi$ pair,
shown in Fig. 3, is obtained; there is a
strong  $K^*(892)$ signal.
 The distribution is fitted with a background polynomial plus a $P$-wave
relativistic
Breit-Wigner function, with a width $$\Gamma = \Gamma_0
{{m_0}\over{m}}{{1+r^2p_0^2}\over{1+r^2p^2}}\bigg\lbrack{{p}\over{p_0}}\bigg\rbrack^3,$$
where $m$ is the mass of the $K\pi$ system, $p$ is the momentum of
kaon in the $K\pi$ system, $\Gamma_0$ is the width of the
resonance, $m_0$ is the mass of the resonance, $p_0$ is $p$ evaluated at the resonance mass, $r$ is the
interaction radius, and $\displaystyle
{{1+r^2p_0^2}\over{1+r^2p^2}}$ represents
the contribution of the barrier factor.
The fit of Fig. 3 gives an $r$ value of $(3.4\pm 2.6)$ GeV$^{-1}$ with
a large
error due to the low statistics. Therefore, in
later analysis (mainly in the efficiency calculation), we use the
value $(3.4\pm0.6\pm0.3)$ GeV$^{-1}$ measured by the $K^-\pi^+$
scattering experiment \cite{ASTON} for $r$.

%\vspace*{5pt}
\begin{figure}[htbp]\begin{center}
\epsfxsize=7.25cm\epsffile{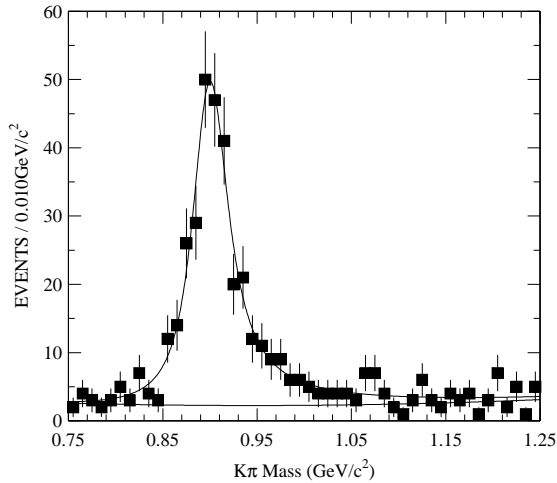}
\label{cpkp2}
\caption{$K\pi$ invariant mass spectrum recoiling against $K^*(892)$~(0.836 GeV $<
m_{K\pi} < 0.956 $ GeV) for events in the $\chi_{cJ}$ mass region
(some events appear twice), where the curves are the Breit-Wigner function and
background polynomial  described in the text.}
\end{center}
\end{figure}

In this paper, the number of
$K^*(892)^0\bar K^*(892)^0$ events and the corresponding background are
estimated from the scatter plot of $K^-\pi^+$ versus $K^+\pi^-$
invariant masses, as shown in Fig. 4.
%This method gives more accurate
%determinations of the $K^*(892)$ signal and background. 
The signal region is shown as a square box (solid line) at $(0.896,
0.896)$ GeV with the width of 60 MeV. From a Monte Carlo study, a
large
 background comes from $\psi(2S)\to \gamma
\chi_{cJ}\to \gamma K_1(1270)K$ (or $K_1(1400)K$) which decays to
$\gamma \pi^+\pi^-K^+K^-$ final states via $K_1\to K^*(892)\pi$
intermediate decay.  This background shows up as the horizontal and
vertical bands at $m(K^*(892))$ in the $m(K^-\pi^+)$ versus $m(K^+
\pi^-)$ scatter plots of Fig. 2.
Hence, backgrounds are estimated from sideband boxes, which are
taken 60 MeV away from the signal box and shown as four dashed-line
boxes in Fig. 4.  Background in the horizontal or vertical
sideband boxes is twice that in the signal region.

\begin{figure}[htbp]
\begin{center}
\epsfxsize=6.25cm\epsffile{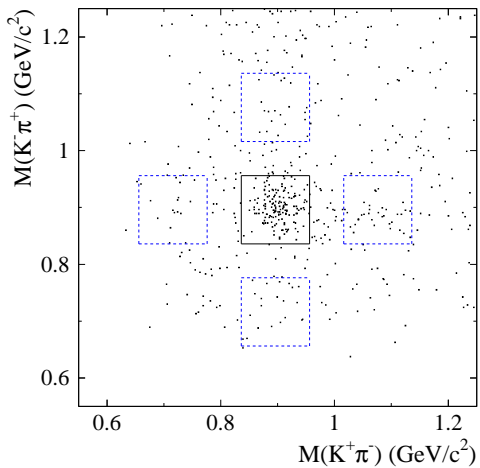}
%\hspace*{0.5cm}
%\epsfxsize=5.0cm\epsffile{mksic-psip.epsi}
\label{cpkp3}
\caption{Definition of signal and  sideband regions.}
\end{center}
\end{figure}

Figure 5 shows the
mass distribution of the $K^*(892)^0\bar K^*(892)^0$ candidate events and the
corresponding background.  There are 154 and 46 (92/2) events obtained
from the signal box and the dashed-line boxes within 3.20 GeV - 3.70~GeV, respectively.

\begin{figure}[htbp]
\begin{center}
\epsfxsize=6.25cm\epsffile{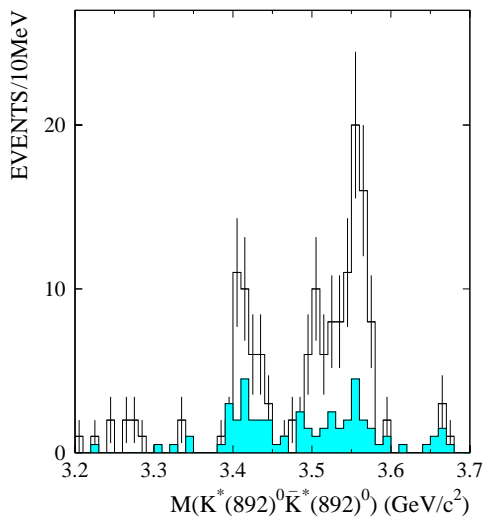}
%\hspace*{0.5cm}
%\epsfxsize=5.250cm\epsffile{mksic-psip.epsi}
\label{cpkp4}
%\vspace*{-5pt}
\caption{The $K^*(892)^0\bar K^*(892)^0$ invariant mass spectrum for
events in the signal region of Fig. 4. The
shaded histogram indicates the distribution of the background
estimated for events in the sideband regions of Fig. 4.}
\end{center}
\end{figure}

\subsection{Fit of the mass spectrum}

After sideband subtraction, the $K^*(892)^0\bar K^*(892)^0$ mass
spectrum between 3.20 and 3.70 GeV is fitted using a $\chi^2$ method with
three Breit-Wigner functions folded with Gaussian resolutions, where
the  mass resolutions are fixed at their Monte Carlo predicted values
[$(12.2\pm0.4)$ MeV, $(12.3\pm0.3)$ MeV and $(12.2\pm0.3)$ MeV for
$\chi_{c0}$, $\chi_{c1}$ and $\chi_{c2}$, respectively] and the widths
of the three $\chi_{cJ}$ states are set at their world average values
\cite{PDG}.
A
$\chi^2$ probability of 70\% is obtained, indicating a reliable
fit. The number of events determined from the fit are
$26.1\pm5.8$, $26.9\pm5.4$ and $55.1\pm6.3$ for $\chi_{c0}$,
$\chi_{c1}$, and $\chi_{c2}$, respectively. The statistical
significances
of the three states are $4.2\sigma$, $4.3\sigma$ and $7.5\sigma$,
calculated
from $\sqrt{\Delta\chi^2}$, where
$\Delta\chi^2$ is the difference between the $\chi^2$ values of
the fits determined with and without the signal function. Fig.~6
shows the fit result, and the fitted masses are
$3416.2\pm3.6$ MeV,
$3507.8\pm3.6$ MeV and $3553.6\pm 1.8$ MeV for
$\chi_{c0}$, $\chi_{c1}$ and $\chi_{c2}$, respectively, in agreement
with the world average values~\cite{PDG}.

\begin{figure}[htbp]
\begin{center}
\epsfxsize=8.5cm\epsffile{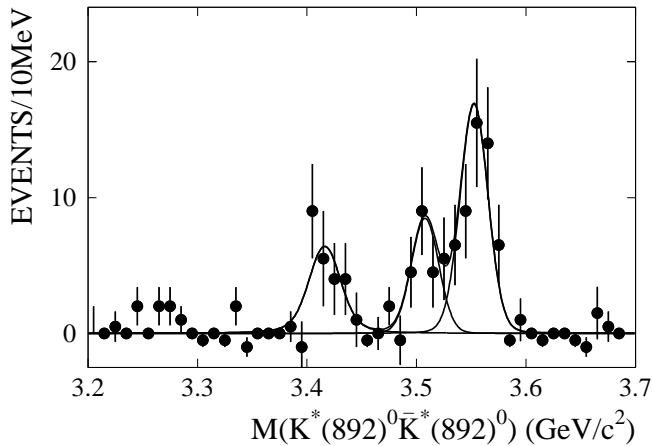}
%\hspace*{0.5cm}
%\epsfxsize=5.250cm\epsffile{mksic-psip.epsi}
\label{cpkp5}
\vspace*{-5pt}
\caption{The $K^*(892)^0\bar K^*(892)^0$ invariant mass spectrum
fitted with three resolution smeared Breit-Wigner functions as
described in the text.}
\end{center}
\end{figure}

A Monte Carlo simulation is used to determine the detection
efficiency. The angular distribution of the photon emitted
in $\psi(2S)\to\gamma\chi_{c0}$ is taken into
account \cite{E1}. The $K^*(892)$ is generated as a $P$-wave
relativistic Breit-Wigner with $r$ as 3.4
GeV$^{-1}$ \cite{ASTON}.  For each case, 50,000 Monte Carlo events are simulated,
and the efficiencies are estimated to be $\epsilon_{\chi_{c0}} =
(3.15\pm0.09)\%,~\epsilon_{\chi_{c1}} = (3.25\pm0.09)\%$, and
$\epsilon_{\chi_{c2}} = (2.96\pm0.08)\%$, where the error is the
statistical error of the Monte Carlo sample. Note that for the
efficiency estimation, the events in the four sideband boxes are
subtracted from the events in the signal region of the scatter plot,
similar to the treatment of data.

The branching fraction of $\psi(2S)\to\gamma\chi_{cJ},~\chi_{cJ}\to
K^*(892)^0\bar 
K^*(892)^0$ is calculated using
\begin{displaymath}
   {\cal B}(\chi_{cJ}\rightarrow \gamma\chi_{cJ}\to\gamma K^{*}(892)^0\bar
K^{*}(892)^0)=\frac{n^{obs}/(\varepsilon\cdot
   f^2)}
{N_{\psi(2S)}},
\end{displaymath}
where the factor $f$ is the branching fraction of $K^*(892)^0$ 
to the charged $K\pi$ mode, which is taken as $\displaystyle 
{2}\displaystyle\over\displaystyle{3}$.

Using the numbers obtained
above and the total number of $\psi(2S)$ events,
$14.0~(1.00\pm0.04)\times 10^6$ \cite{moxh}, we determine the following
branching
fractions 
$${\cal B}(\psi(2S)\to\gamma\chi_{c0}\to\gamma K^{*}(892)^0\bar
K^*(892)^0) = (1.33\pm0.30)\times 10^{-4},$$
$${\cal B}(\psi(2S)\to\gamma\chi_{c1}\to\gamma K^*(892)^0\bar
K^*(892)^0) = (1.33\pm0.27)\times 10^{-4},$$
$${\cal B}(\psi(2S)\to\gamma\chi_{c2}\to\gamma K^*(892)^0\bar
K^*(892)^0) = (2.99\pm0.35)\times 10^{-4},$$
where the errors are statistical only.

\subsection{Systematic errors}

The systematic errors in the branching fraction measurement associated
with the efficiency are determined by comparing $\psi(2S)$ data and
Monte Carlo simulation for very clean decay channels, such as
$\psi(2S)\to\pi^+\pi^-J/\psi$, which allows the determination of
systematic errors associated with the MDC tracking, kinematic fitting,
particle identification, and
efficiency of
the photon ID \cite{CZ}.
Other sources of systematic error come from
the uncertainties in the number of $\psi(2S)$ events \cite{moxh},
the efficiency estimation using simulated data, the background, the
$\chi_{cJ}$ and $K^*(892)^0$ mass resolutions, the binning and fit range, etc.

\subsubsection{Efficiency estimation}
As mentioned above, we use the measurement of Ref.~\cite{ASTON},
$(3.4\pm0.6\pm0.3)$ GeV$^{-1}$ for $r$ in the $P$-wave relativistic
Breit-Wigner parameterization in the Monte Carlo
simulation. We also use $r$ varied by one sigma
to 2.73 GeV$^{-1}$ and 4.07
GeV$^{-1}$
to determine the change in the detection efficiency.
For $r = 2.73$ GeV$^{-1}$, the efficiencies of the
$\chi_{c0}$,~$\chi_{c1}$,
and $\chi_{c2}$ become $2.90\%$, $2.99\%$ and
$2.74\%$, and for $r = 4.07$ GeV$^{-1}$, the
corresponding efficiencies are $3.37\%$, $3.52\%$,
and
$3.16\%$, respectively. The largest changes are about 7.9\%,
8.3\% and 7.4\% for $\chi_{c0}$, $\chi_{c1}$
and $\chi_{c2}$.

\subsubsection{Background subtraction}
In Section IV A, the backgrounds are estimated using the sidebands
shown as the four dashed-line boxes in Fig. 4. Moving the sideband
boxes 20 MeV away from or closer to the signal region, or varying the
background number by one standard deviation, the largest changes of the
branching fractions for the $\chi_{c0}$, $\chi_{c1}$ and $\chi_{c2}$
are about 7.4\%, 5.0\% and 5.5\%, respectively, obtained by re-fitting
the $K^*(892)^0 \bar K^*(892)^0$ mass spectrum and reestimating the efficiency.

\subsubsection{$\chi_{cJ}$ and $K^*(892)^0$ mass
resolutions} Differences between data and Monte Carlo for the mass
resolutions of the $\chi_{cJ}$ or $K^*(892)^0$ also give uncertainties
in the determination of the branching fractions.  The maximum possible
difference for $\chi_{cJ}$ is about 1 MeV.  Such a change results in
about 4.5\%, 2.5\% and 2.0\% variations in the fitted number of $\chi_{c0}$,
$\chi_{c1}$, and $\chi_{c2}$ events. If we change the $K^*(892)^0$
window to [0.836 + 0.002,~0.956 - 0.002] GeV and [0.836 - 0.002,~0.956
+ 0.002] GeV, the efficiency variations of the $\chi_{c0}$,
$\chi_{c1}$ and $\chi_{c2}$ are 1.5\%, 2.5\% and 2.4\%,
respectively. By varying the width of $\chi_{c0}$ by 1$\sigma$, 0.8
MeV, there is almost no change in the final fit result.  We use total
systematic errors of 5\%, 3.5\% and 3.5\% for this uncertainty.

\subsubsection{Binning and fit range}
Using different binning and fit ranges for the
$K^{*}(892)^0\bar
K^{*}(892)^0$ mass spectrum fit yields errors of
about 4\%, 2\% and 2\% for $\chi_{c0}$, $\chi_{c1}$ and
$\chi_{c2}$, respectively.

\begin{table}[htbp]
\begin{center}
\caption{Summary of systematic errors in the branching fraction
calculation of ${\cal B}(\psi(2S)\to\gamma\chi_{cJ}\to\gamma
K^{*}(892)^0\bar
K^{*}(892)^0)$.}
\begin{tabular}{lccc}
\hline
\hline
Source~~~~~~~~~~~~~~~~&~~~~~~~~~$\chi_{c0}$~~~&~~~~~~$\chi_{c1}$~~~~~~&~~~$\chi_{c2}$\\
\hline
MDC tracking&~~~~~~~~~8\%&8\%&8\%\\
Kinematic fit&~~~~~~~~~6\%&6\%&6\%\\
Particle identification&~~~~~~~~~5\%&5\%&5\%\\
Photon ID efficiency&~~~~~~~~~2\%&2\%&2\%\\
$\psi(2S)$ number&~~~~~~~~~4\%&4\%&4\%\\
Efficiency estimation&~~~~~~~~7.9\%&8.3\%&7.4\%\\
Background&~~~~~~~~7.3\%&5.0\%&5.5\%\\
Mass
resolutions&~~~~~~~~5\%&3.5\%&3.5\%\\
Binning and fit range&~~~~~~~~4\%&2\%&2\%\\
\hline
Total systematic error &~~~~~~~~17.4\%&16.0\%&15.7\%\\
\hline\hline

\end{tabular}
\end{center}
\end{table}

\vspace*{15pt}
The systematic errors from all sources are listed in Table I, as are
the total errors of $17.4\%$, 16.0\% and 15.7\% for $\chi_{c0}$,
$\chi_{c1}$ and
$\chi_{c2}$, respectively, obtained by adding them in quadrature. The resulting branching fractions are
$${\cal B}(\psi(2S)\to\gamma\chi_{c0}\to\gamma K^*(892)^0\bar
K^*(892)^0) = (1.33\pm0.30\pm0.23)\times 10^{-4},$$
$${\cal B}(\psi(2S)\to\gamma\chi_{c1}\to\gamma K^*(892)^0\bar
K^*(892)^0) = (1.33\pm0.27\pm0.21)\times 10^{-4},$$
$${\cal B}(\psi(2S)\to\gamma\chi_{c2}\to\gamma K^*(892)^0\bar
K^*(892)^0) = (2.99\pm0.35\pm0.47)\times 10^{-4},$$
and
with the PDG world average values of
$\psi(2S)\to\gamma\chi_{cJ}$  \cite{PDG}, we get the following
branching fractions
$${\cal B}(\chi_{c0}\to K^*(892)^0\bar
K^*(892)^0) = (1.55\pm0.35\pm0.30)\times 10^{-3},$$
$${\cal B}(\chi_{c1}\to K^*(892)^0\bar
K^*(892)^0) = (1.58\pm0.32\pm0.29)\times 10^{-3},$$
$${\cal B}(\chi_{c2}\to K^*(892)^0\bar
K^*(892)^0) = (4.67\pm0.55\pm0.85)\times 10^{-3},$$
where the first error is statistical and the second is
systematic. The numbers used and results are summarized in Table II.

\begin{table*} \centering
\caption{Summary of numbers used in the branching fraction calculation
and branching fraction results.}
\begin{ruledtabular}
\begin{tabular}{lccc}  
Quantity~~~~~&$\chi_{c0}$~~~&~~~~~~$\chi_{c1}$~~~~~~&~~~$\chi\
_{c2}$\\
\hline
$n^{obs}$&$26.1\pm5.8$&$26.9\pm5.4$&$55.1\pm6.3$\\
%\hline
$\epsilon~(\%)$&$3.15\pm0.09$&$3.25\pm0.09$&$2.96\pm0.08$\\
$N_{\psi(2S)}~(10^6)$ \cite{moxh}&$14.0\pm 0.6$ &$14.0\pm 0.6$ &$14.0\pm 0.6$ \\
$f$&$\displaystyle {2}\displaystyle\over\displaystyle{3}$&$\displaystyle {2}\displaystyle\over\displaystyle{3}$&$\displaystyle {2}\displaystyle\over\displaystyle{3}$\\
${\cal
B}(\psi(2S)\to\gamma\chi_{cJ})$ (\%) \cite{PDG}&$8.6\pm0.7$&$8.4\pm0.8$&$6.4\pm0.6$\\
${\cal
B}(\chi_{cJ}\to K^*(892)^0\bar
K^*(892)^0)$ $(10^{-3})$&$1.55\pm0.35\pm0.30$&$1.58\pm0.32\pm0.29$&$4.67\pm0.55\pm0.85$\\
%~~~~~~~~~~~~~~\,~$(10^{-3})$&(1.53\pm0.35\pm0.30)\times 10^{-3}&$1.58\pm0.32\pm0.28$&$4.40\pm0.52\pm0.79$\\
\end{tabular} \\
\end{ruledtabular}
\end{table*}

\section{Summary}
In summary, $K^*(892)^0\bar K^*(892)^0$ signals from $\chi_{cJ}~(J=0,1,2)$ decays are observed for  the first time
using a sample of 14 million $\psi(2S)$ events accumulated at the
BES\,II detector.
The branching fractions are determined to be
${\cal B}(\chi_{c0}\to
K^*(892)^0$
$\bar K^*(892)^0) = (1.55 \pm 0.35~(\mbox{stat})
\pm 0.30~(\mbox{syst}))\times
10^{-3}$, ${\cal B}(\chi_{c1}\to
K^*(892)^0$
$\bar K^*(892)^0)$
$ = (1.58 \pm 0.32~(\mbox{stat})
\pm 0.29~(\mbox{syst}) )\times
10^{-3}$, and ${\cal B}(\chi_{c2}\to
K^*(892)^0\bar K^*(892)^0) = (4.67 \pm 0.55~(\mbox{stat})
\pm 0.85~(\mbox{syst}) )\times
10^{-3}$; and the significances of the $K^*(892)^0\bar
K^*(892)^0$ signals are about 4.2$\sigma$, 4.3$\sigma$ and 7.5$\sigma$ for
the
$\chi_{c0,1,2}$ decays, respectively.
This will be helpful in understanding the nature of $\chi_c$ states.

\section{Acknowledgements}

   The BES collaboration thanks the staff of BEPC for their hard
   efforts.
This work is supported in part by the National Natural Science
   Foundation
of China under contracts Nos. 19991480, 10225524, 10225525, the Chinese
   Academy
of Sciences under contract No. KJ 95T-03, the 100 Talents Program of
   CAS
under Contract Nos. U-11, U-24, U-25, and the Knowledge Innovation
   Project of
CAS under Contract Nos. U-602, U-34 (IHEP); by the National Natural
   Science
Foundation of China under Contract No. 10175060 (USTC), and
   No. 10225522 (Tsinghua University); and by the
   Department
of Energy under Contract No.
DE-FG03-94ER40833 (U Hawaii).

%\newpage %Just because of unusual number of tables stacked at end
%\bibliography{gkk}% Produces the bibliography via BibTeX.

\end{document}